\documentclass[twocolumn]{aastex631}

\makeatletter
\newcommand\footnoteref[1]{\protected@xdef\@thefnmark{\ref{#1}}\@footnotemark}
\makeatother

\shortauthors{Kumar et al.}
\graphicspath{{./}{figures/}}

\begin{document}

\title{Resolving the bow shock and tail of the cannonball pulsar PSR J0002+6216}

\author{P. Kumar}
\affiliation{Department of Physics and Astronomy, University of New Mexico, 210 Yale Blvd NE, Albuquerque, NM 87106, USA}

\author{F. K. Schinzel}
\altaffiliation{An Adjunct Professor at the University of New Mexico.}
\affiliation{National Radio Astronomy Observatory, P. O. Box O, Socorro, NM 87801, USA }

\author{G. B. Taylor}
\affiliation{Department of Physics and Astronomy, University of New Mexico, 210 Yale Blvd NE, Albuquerque, NM 87106, USA}

\author{M. Kerr}
\affiliation{Space Science Division, Naval Research Laboratory, Washington, DC 20375–5352, USA}

\author{D. Castro}
\affiliation{NASA Goddard Space Flight Center, Greenbelt, MD 20771, USA}
\affiliation{Harvard-Smithsonian Center for Astrophysics, Cambridge, MA 02138, USA}

\author{U. Rau}
\affiliation{National Radio Astronomy Observatory, P. O. Box O, Socorro, NM 87801, USA }

\author{S. Bhatnagar}
\affiliation{National Radio Astronomy Observatory, P. O. Box O, Socorro, NM 87801, USA }



\begin{abstract}

We present X-ray and radio observations of the recently-discovered bow
shock pulsar wind nebula associated with PSR~J0002+6216, characterizing
the PWN morphology, which was unresolved in previous studies. The
multi-frequency, multi-epoch Very Large Array radio observations reveal
a cometary tail trailing the pulsar and extending up to 5.3$\arcmin$,
with multiple kinks along the emission. The presented radio continuum
images from multi-configuration broadband VLA observations are one of
the first results from the application of multi-term
multi-frequency synthesis deconvolution in combination with the
awproject gridder implemented in the Common Astronomy Software
Applications package (CASA). The X-ray emission observed with Chandra
extends to only 21$\arcsec$, fades quickly, and has some hot spots
present along the extended radio emission. These kinks could indicate
the presence of density variation in the local ISM or turbulence. The
bow shock standoff distance estimates a small bow shock region with a
size 0.003--0.009 pc, consistent with the pulsar spin-down power
of $\dot{E}$=1.51x$10^{35}$ \,ergs s$^{-1}$ estimated from timing. The high
resolution radio image reveals the presence of an asymmetry in the bow
shock region which is also present in the X-ray image. The broadband
radio image shows an unusually steep spectrum along with a flat-spectrum
sheath, which could indicate varying opacity or energy injection into
the region. Spatially-resolved X-ray spectra provide marginal evidence
of synchrotron cooling along the extended tail. Our analysis of
the X-ray data also shows that this pulsar has a low spin-down power and one of the lowest X-ray efficiencies observed in
these objects.
\end{abstract}



\section{Introduction} \label{sec:intro}
Studying the dynamics inside complex supernova environments is challenging. In many cases, the supernova explosion of a massive star leaves a neutron star, and this rapidly spinning collapsed remnant is sometimes detected as a pulsar.  The pulsar loses the bulk of its rotational energy to Poynting flux and a relativistic wind of electrons and positrons accelerated in the the magnetosphere \citep[``injected particles'',][]{KennelCoroniti}. This radiatively cold outflow of particles is shocked upon interaction with the ambient medium, giving rise to a pulsar wind nebula (PWN). Understanding the evolution of the PWN can help constrain the wind properties, along with the progenitor, the local ISM, and the neutron star \citep[see, e.g.,][]{slanepwn,gelfand,vandermhdpwn}. Multi-wavelength observations of PWNe can provide constraints on the geometry of the central engine, density, and composition of the ambient medium, and the spectrum and evolution of injected particles into the surrounding medium \citep[see e.g.,][]{kargaltsevklinger,bucchiantini2018}.\\
At the same time, the evolution of the composite supernova remnant is driven by the evolution of the PWN, where the majority of these remnants retain a spherical shape \citep[see, e.g.,][]{slanepwn}. But the occasional deviation from spherical symmetry is important to understand, arising from complex shapes formed by relic PWN during late evolution as well as high natal kick velocity imparted to some neutron stars by the supernova explosion \citep[][]{hobbs}. Initially the PWN is embedded inside the supernova remnant and its properties depend on e.g. the initial field strength and spin-down timescale of the such neutron star, the surrounding ISM, and the pulsar wind and the supernova explosion energy \citep[see e.g.,][]{Reynolds}, which are difficult to measure directly from the neutron star but can be inferred from the observations of PWNe. These properties are essential to understand the mechanism of core-collapse supernova and neutron star formation as well as the formation and evolution of the progenitor.\\
The size and morphology of a PWN can vary depending on the supernova energy and the properties of the ambient medium. It is often difficult to disentangle these different components, especially in situations where a pulsar wind nebula stays embedded within its supernova remnant shell. Although, in rare cases, the pulsar can get a significant `kick' from the initial explosion and as the supernova decelerates with time, the pulsar may escape its surrounding supernova remnant (SNR) to interact with the surrounding ISM. Only a handful of such systems have been identified e.g. the Mouse and the Mushroom PWN \citep[][]{kargalchandra}.  Even fewer have a robust measurement of proper motion and distance to claim a pulsar velocity (V$_{\mathrm{PSR}}$) in excess of 1000 km s$^{-1}$ \citep[][]{hobbs}. In these rare cases, a bow shock is formed by the supersonic motion of the pulsar through the ISM. As the ram pressure of the motion confines the PWN, the injected shocked particles and magnetic field are swept backward, resulting in broadband emission from a non-thermal synchrotron tail \citep[][]{zhang}. If the ISM is sufficiently neutral, an H$\alpha$ bow shock emission can also be visible due to the ionization of the ambient medium by the pulsar \citep[see, e.g., ][]{guitar}. These bow shock PWN have a simple geometry \citep[][]{morlino}, where there is a clean separation of different components created by the newly shocked winds and the synchrotron flow carried downstream by the ram pressure \citep[][]{bykov}. This makes them ideal systems for studying the particle acceleration, morphology, and evolution of pulsar winds. However, in reality these bow shock PWNe have complex morphologies \citep[][]{kargaltsevklinger}, which may be due to the different ISM conditions, mass loading, geometry, and field orientation, among other possibilities. This sometimes results in jet-like structures emanating from the bow shock \citep[see e.g., ][]{guitar,devries}. Even with these complexities, bow shock PWN are essential for studying the evolution and formation of the composite system. Moreover, the discovery of $\gamma$-ray halos around PWNe make them important to understand the positron excess seen in the cosmic ray spectrum and the search for dark matter \citep[see, e.g., ][]{Hooper}.\\
\vspace{-0.0cm}
The serendipitous discovery of a bow-shock PWN nebula \citep[][]{frank} towards the 115\,ms $\gamma$-ray pulsar PSR\,J0002+6216 provides one such system to study the properties of the bow shock region and the particle evolution in the tail region. From PWN properties and measured proper motion of the pulsar from timing and other methods, as well as the geometry of the PWN tail pointing towards the geometric center of SNR CTB1, they argued that the pulsar was born in the same core-collapse supernova which gave rise to SNR CTB1. Based on PWN properties and timing data from \textit{Fermi} \citep[][]{frank}, the pulsar has a velocity in excess of 1000 km s$^{-1}$, making it one of the fastest objects in its category, where the distribution peaks around a few hundred km s$^{-1}$. Using 1--2\,GHz radio observations, the authors show the presence of a narrow, uniformly bright tail, which is marginally resolved, with spectral index, $\alpha$ = -0.98$\pm$0.31, which is atypical of the range for PWNe of -0.5 to 0 found by \citet{weilersramek}. Estimation of the magnetic field B, synchrotron lifetime t$_\mathrm{{syn}}$, and relativistic gas pressure in the tail hints towards a small shock standoff distance and a large Mach number (M$\sim$200) for the PWN \citep[][]{frank}. Here we present follow-up broadband radio continuum and Chandra X-ray observations to test these claims. These spectrally resolved X-ray and radio observations can be used to look for evidence of in-situ particle acceleration, synchrotron cooling and to measure the size of the X-ray nebula which relates to the ratio of the age of the nebula to the synchrotron lifetime of the electrons. Similarly, our high resolution radio observations can confirm the unusual radio spectrum of the nebula, along with confirming the orientation as well as measuring the size of the PWN bow shock.  \\
\vspace{-0.0cm}
In Section \ref{sec:observe}, we describe the observation and data reduction procedures for Karl G. Jansky Very Large Array (VLA) and Chandra observations. Section \ref{sec:results} provides the main results and analysis of the observed radio and X-ray properties, Section \ref{sec:discuss} discusses the results in the context of our current understanding of pulsar wind nebula physics, with Section \ref{sec:summary} briefly summarizing the key findings presented here. 


\section{Observations and Data reduction} \label{sec:observe}
\subsection{Radio Observations}

We observed a field towards PSR~J0002+6216 with the Karl G. Jansky VLA \citep[][]{perley} at four different epochs using the 4--8\,GHz (C band) and 8--12\,GHz (X band) receivers and with the standard Wideband Interferometric Digital Architecture (WIDAR) correlator setup for continuum observing. The observations were carried out in 4 different epochs under NRAO proposal IDs VLA/18B-397, SL0115, and 21B-334. The details of the observational setup for the C and X band observations are described in Table \ref{tab:3} and \ref{tab:4} respectively. In all of the observations above, the radio source J0102+5824 was used as a phase calibrator, while 3C 48 (J0137+331) was used as both the bandpass and flux density calibrator. Part of the data were observed allowing for full polarization calibration. The observation on February 17 was significantly affected by bad weather, specifically accumulated snow in the telescope dishes causing signal attenuation.\\

\begin{deluxetable*}{ccccccc}[htbp!]
\tablecolumns{7}
\tabletypesize{\small}
\tablewidth{0pt}
\tablecaption{4--8\,GHz C-band Observation Setup \label{tab:3}}
\tablehead{
    \colhead{Epoch} & \colhead{Proposal ID} & \colhead{VLA Configuration} & \colhead{Spectral Windows} & \colhead{Channels} & \colhead{Integration Time} & \colhead{Time On Source} \\
    \colhead{ } & \colhead{ } & \colhead{ } & \colhead{ } & \colhead{(MHz)} & \colhead{(s)} & \colhead{(min)}
}
\startdata
17 Feb 2019 & 18B-397 & B & 32 & $128\times1$ & 2& 47 \\
24 Feb 2019 & 18B-397 & B & 16 & $128\times2$ & 2& 47 \\
25 Jan 2020 & SL0115 &  D & 32 & $64\times2$ & 2& 103 \\
\enddata
\tablecomments{Observational setup for all the 4--8\,GHz (C band) observations. All observations were done with 4\,GHz of total bandwidth centered on 6\,GHz.}
\end{deluxetable*}

\begin{deluxetable*}{ccccccc}[htbp!]
\tablecolumns{7}
\tabletypesize{\small}
\tablewidth{0pt}
\tablecaption{ 8--12\,GHz C X-band Observation Setup\label{tab:4}}
\tablehead{
    \colhead{Epoch} & \colhead{Proposal ID} &
    \colhead{VLA Configuration} &
    \colhead{Spectral Windows} & \colhead{Channels} & \colhead{Integration Time} & \colhead{Time On Source} \\
    \colhead{} & \colhead{} &
    \colhead{} &
    \colhead{} & \colhead{(MHz)} & \colhead{(s)} & \colhead{(min)}
}
\startdata
17 Feb 2019 & 18B-397 & B & 32 & $128\times1$ & 2& 144 \\
24 Feb 2019 & 18B-397 & B & 16 & $128\times2$ & 2& 144 \\
25 Aug 2021 & 21B-334 & C & 32 & $64\times2$ & 2& 140 \\
\enddata

\tablecomments{Observational setup for all the 8--12\,GHz (X band) observations. All observations were done with 4\,GHz of total bandwidth centered on 10\,GHz.}
\end{deluxetable*}

\begin{deluxetable*}{cccccccc}[htbp!]
\tablecolumns{8}
\tabletypesize{\small}
\tablewidth{0pt}
\tablecaption{Imaging Parameters and Final Image Properties \label{tab:5}}
\tablehead{
    \colhead{Image} & \colhead{Frequencies} &
    \colhead{VLA Configuration} &
    \colhead{Brigg's weighting} & \colhead{Synthesized Beam} & \colhead{Resolution} & \colhead{Spatial Size} & \colhead{Noise rms}\\
    \colhead{No.} & \colhead{(GHz)} &
    \colhead{} &
    \colhead{} & \colhead{} & \colhead{} & \colhead{} &\colhead{($\mu$Jy\, beam$^{-1}$)}
}
\startdata
1 & 4-12 & all available & 0.5 & 1.78$\arcsec$ x 1.03$\arcsec$ & 0.45$\arcsec$ & 16.2$\arcmin\times16.2\arcmin$ & 1.08 \\
2 & 8-12 & B (Epoch 1) & 0.5 & 1.13$\arcsec$ x 0.56$\arcsec$ & 0.075$\arcsec$ & 9$\arcmin\times9\arcmin$ & 2.08 \\
\enddata
\tablecomments{CASA imaging parameters used and the final image properties of the two radio continuum images shown here. Figure \ref{fig:1}, shows a zoom in on the bow shock and tail of emission, and Figure \ref{fig:2} shows the tip of the bow shock region, hence the image area is not the same as mentioned in the table.}
\end{deluxetable*}

\vspace{-3cm}

The correlated visibilities for all observing frequencies were calibrated using CASA 6.2.1.7 together with the automated VLA data reduction pipeline \citep[]{casa}. For the observation on February 17th only, manual data reduction was performed by transferring flux densities derived for the complex gain calibrator on February 24th to compensate for effects from snow accumulation in the antennas. The calibrated measurement sets had their statistical weights reinitialized prior to imaging according to the integration time and bandwidth to account for any additional data flagging that was performed. Imaging was done in CASA 6.2.1.7 using the wideband AWProjection gridding algorithm in combination with multi-term and multi-frequency synthesis \citep[]{Raucornwell,bhatnagar,Raubhatnagar}, and presents one of the first imaging results from the application of this broadband imaging algorithm. For imaging, we used the ``Briggs" weighting scheme \citep[][]{danbriggs}, which is a trade off between noise in the image for resolution. No w-projection correction was applied for imaging. We tried imaging with 8 w-projection planes and there was no significant improvement in image features or noise. We created four separate radio continuum images which show 1) radio emission from the pulsar, 2) the bow shock, 3) the extended tail behind it, 4) and the asymmetry in the bow shock region. These include three from the combined data and one from a single epoch of longest baseline data and thus highest resolution (Epoch 1: 17 Feb 2019, 8--12\,GHz). Among these four images, those obtained from imaging only the combined C-band data and only the combined X-band data do not show any features not present in the combined broadband 4--12\,GHz image. Hence, we only show the combined broadband 4--12\,GHz image and the highest resolution single epoch 8--12\,GHz image, as pointed out in Table \ref{tab:4}. For the combined data images, the calibrated data were converted to $2$\,MHz spectral channels. All images are centered on R.A./decl. pointing of 00$^h$2$^m$58$^s$.15 and +62$\degr$16$\arcmin$19$\arcsec$.46. The final imaged area, in each case, is well beyond the half power point, by about a factor of 2.\\

\begin{figure*}
    \centering
    \plotone{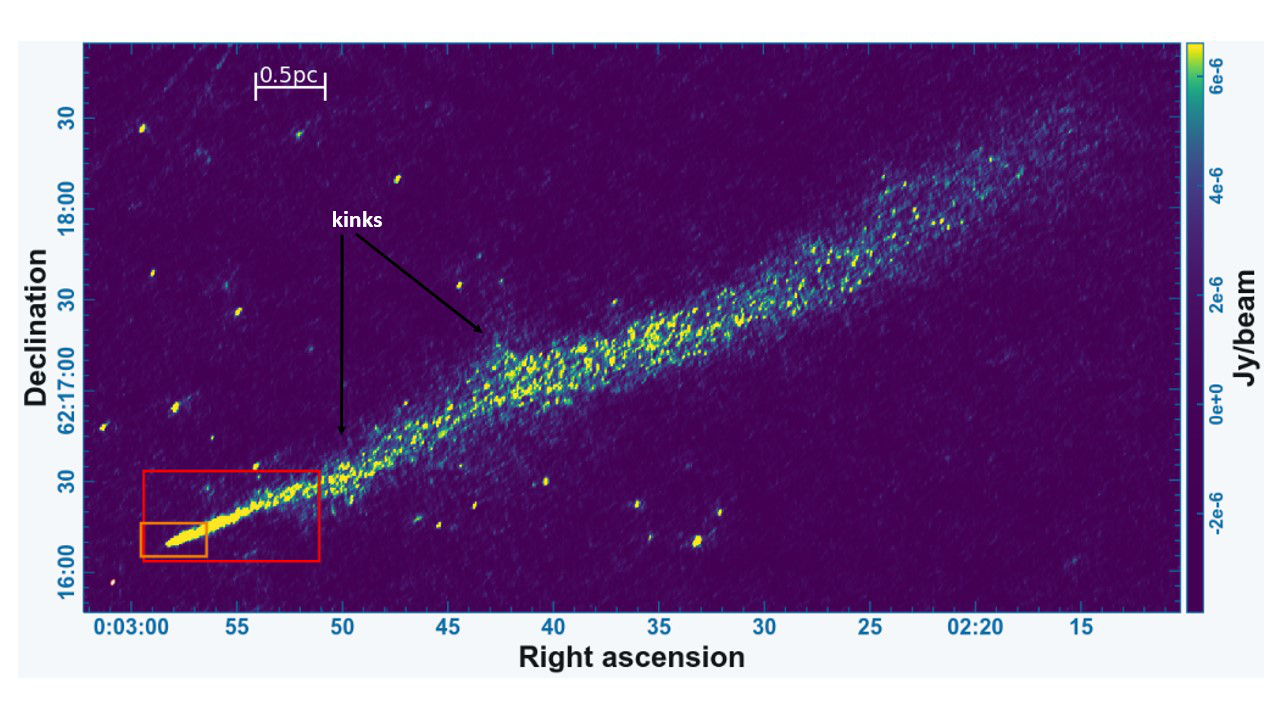}
    \caption{The combined 4--12\,GHz continuum image of the pulsar, zooming in on the bow shock and the tail of emission. It clearly shows a bright extended tail of emission behind the pulsar as it moves through the ISM, along with kinks in the tail. The image properties are listed in Table \ref{tab:5}. The top left marker represents the physical size in the image assuming a source distance of 2\,kpc from \citet{frank}. The brown and red box represent the spatial area covered by the image in Figures \ref{fig:2} and \ref{fig:4} respectively. The red ellipse with white cross in the lower left corner shows the size of the synthesized beam of 1.78$\arcsec$ x 1.03$\arcsec$.}
    \label{fig:1}
\end{figure*}

\vspace{-0.4cm}
Combining all available data, we created a combined C and X band, 4--12\,GHz continuum image, centered at 8\,GHz, with corresponding image properties shown in Table \ref{tab:5}. The noise in the combined broadband image is above the theoretical sensitivity of the VLA, but still lower than a factor of two.  A zoom in on the combined image is shown in Figure \ref{fig:1}. Additionally, a naturally weighted image was also created for the combined C and X band data, to get the best spectral index image of emission.\\

\begin{figure*}
    \centering
    \plotone{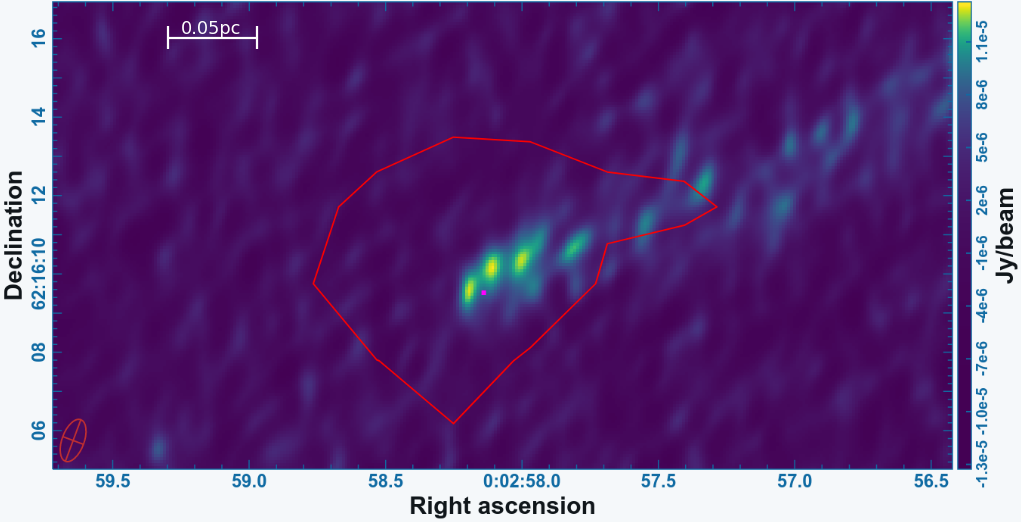}
    \caption{The 8--12\,GHz continuum image, zooming in on the tip of the bow shock region, was created with only the higher resolution B configuration data from the first epoch of observations, with the location of the pulsar shown by the magenta dot and the red contour from the X-ray image in Figure \ref{fig:3} showing the shape of the bow shock. The image clearly shows asymmetric emission inside the bow shock region, misaligned from the location of the pulsar. The image properties are listed in Table \ref{tab:5}, serial number 4. The top left marker represents the physical size in the image assuming a source distance of 2\,kpc from \citet{frank}. The synthesized beam of size 1.13$\arcsec$ x 0.56$\arcsec$ is shown by the red ellipse in lower left corner.}
    \label{fig:2}
\end{figure*}

\vspace{-0.4cm}
Finally, in an effort to resolve bow shock region we show the highest resolution, longest baseline, 8--12\,GHz X band image, to look at the details in the bow shock region. The image properties are listed in Table \ref{tab:5}. The imaged area goes up to the edge of the primary beam, with the image noise about a factor of two higher than the theoretical noise limit of the VLA. An image showing the zoom in on the tip of the bow shock region, covering the same spatial area as the brown box in Figure \ref{fig:1} is shown in Figure \ref{fig:2}.

\subsection{X-ray observations}\label{xrayobs}
\begin{figure*}
    \centering
    \plotone{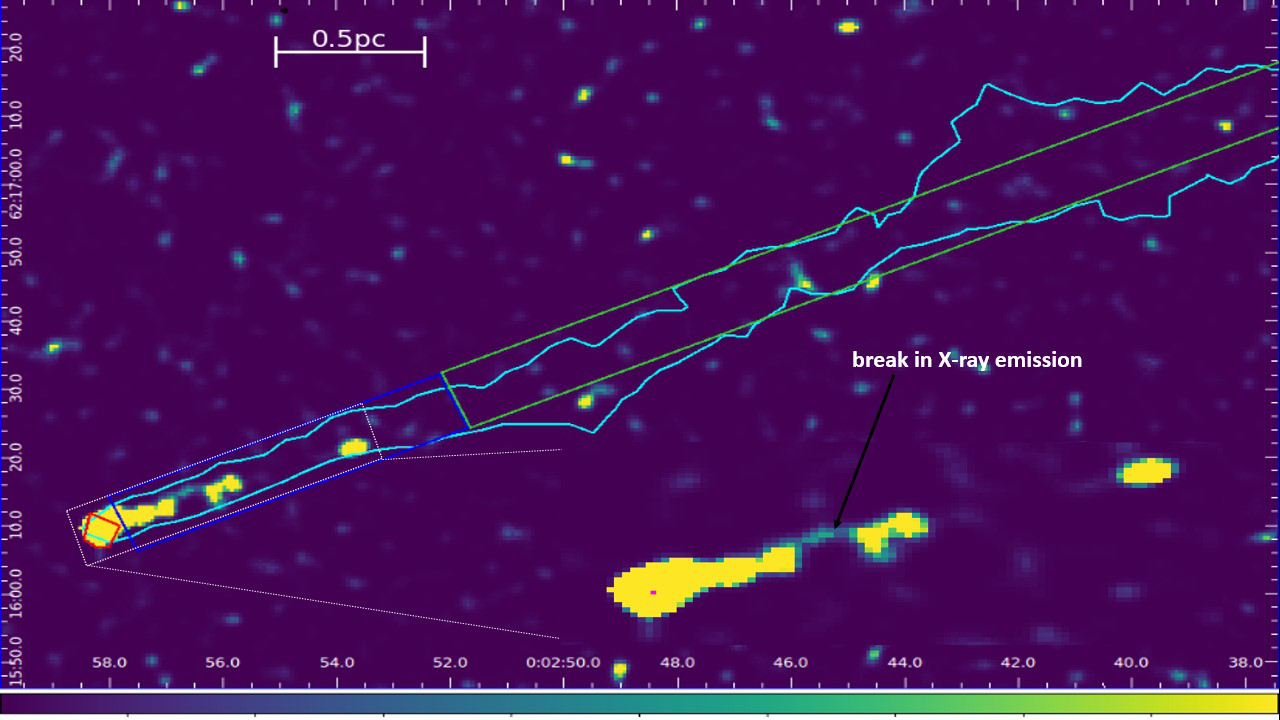}
    \caption{The $0.5 -7.0$\,keV image of the pulsar along with the X-ray tail, taken with the ACIS instrument on Chandra. The three different analysis regions are the red (r${_0}$), blue (r${_1}$), and green (r${_2}$) boxes and the radio contour from the 4--12\,GHz image is marked in cyan. The image is generated from the combined data taken at two epochs totaling 46.3 ks. The image is created in CIAO-ds9. The top left marker represents the physical size in the image assuming a source distance of 2\,kpc from \citet{frank}. The inset shows the zoom in on the X-ray emission from the PWN, with the position of the pulsar marked by the magenta dot from \citet{frank}.}
    \label{fig:3}
\end{figure*}

We carried out X-ray observations of a field towards PSR~J0002+6216 using Chandra X-ray Observatory, with the Advanced CCD Imaging Spectrometer (ACIS-S), in two epochs (Observation ID: 22428 and 23278), with 30.05 and 16.87 ks exposures respectively and analyzed with Chandra Interactive Analysis of Observations (CIAO) software version 4.12 \citep[][]{ciao}. Each dataset was reprocessed with the most recently available calibration applied to it using \textit{chandra\_repro}, which creates new level two event files for each dataset which are used for all further processing and analysis. Initial imaging of each of the individual datasets and registering them to the radio coordinate frame shows a slight astrometric shift between them. So first we correct for this shift for each dataset using the combined 8--12\,GHz X band radio image. This is done by identifying significant point sources in the radio image using PyBDSF \citep[][]{PyBDSFPB} and using \textit{wavedetect} in CIAO for the X-ray images. Then an appropriate correction is applied to the X-ray data using the tools \textit{wcs\_match} and \textit{wcs\_update} in CIAO. This resulted in a translational shift of -1.15 to 1.67 in X and Y pixel positions.\\
Once corrected, separate images were created for each dataset to identify spatially resolved regions on the pulsar and along the tail, for spectral analysis. The images were created using the tool \textit{fluximage}. Additionally, a combined image is created using the tool \textit{merge\_obs} as shown in Figure \ref{fig:3}.\\
\vspace{-0.0cm}
Since there is a slight offset in the roll angle between the two observations we cannot use the combined data from \textit{merge\_obs} for spectral analysis, as this can lead to incorrect response function calculations, which will affect the spectral analysis. Hence, spectral regions at the same spatial locations are extracted from each observation, and then jointly fitted for spectral analysis. For that, we overlay a radio contour from the 8--12\,GHz VLA observations on the combined X-ray image. Then spatially resolved regions are created separating the pulsar and the tail region, using standard tools in ds9 \citep[][]{DS9}, as shown by the red (r${_0}$), blue (r${_1}$) and green (r${_1}$) boxes in Figure \ref{fig:3}, and saved in a CIAO WCS format. These regions are then converted to PHYSICAL format using the tool \textit{dmmakereg} for both the observations using their corresponding broadband flux, which is the $0.5-7.0$\,keV image. Additionally, a background region is also created for each of the two observations.\\
\vspace{-0.0cm}
After obtaining the source regions and the background file, we used the script \textit{specextract} to extract the source spectrum for each of the two observations, for the different regions. The next step involves combining the different spectra from the two observations, which was done using the task \textit{combine\_spectra}. Once we have the combined spectrum of the different regions, we load it into the analysis package \textit{SHERPA} for further analysis.\\
\vspace{-0.0cm}
For fitting the spectrum to the data, we first apply an ignore function to remove any data outside the $0.5-7.0$\,keV energy range. Then we group the counts in bins of 20. These tasks are carried out using \textit{ignore} and \textit{group\_counts} functions respectively. After this, we create an absorption power law model for fitting the source and the background simultaneously, where the number density is supplied from the HI4PI survey catalog \citep[][]{h14pi}, in the direction of the pulsar and is frozen for fitting purposes. The only fitting parameters are the photon index and and the normalization value. After setting the source model, the \textit{fit} function is used to fit the data using \textit{cash} statistics, followed by using \textit{conf} to calculate the 90$\%$ confidence intervals for the fitted parameters. Then we calculate the 90$\%$ bounds on the energy flux for each region using \textit{sample\_energy\_flux} in the energy range of $0.5-8.0$\,keV, followed by using scipy to calculate the bounds from the derived distribution. We extrapolate the fluxes slightly beyond the $0.5-7.0$\,keV data range for consistency with other reported results.

\begin{figure*}
    \centering
    \plotone{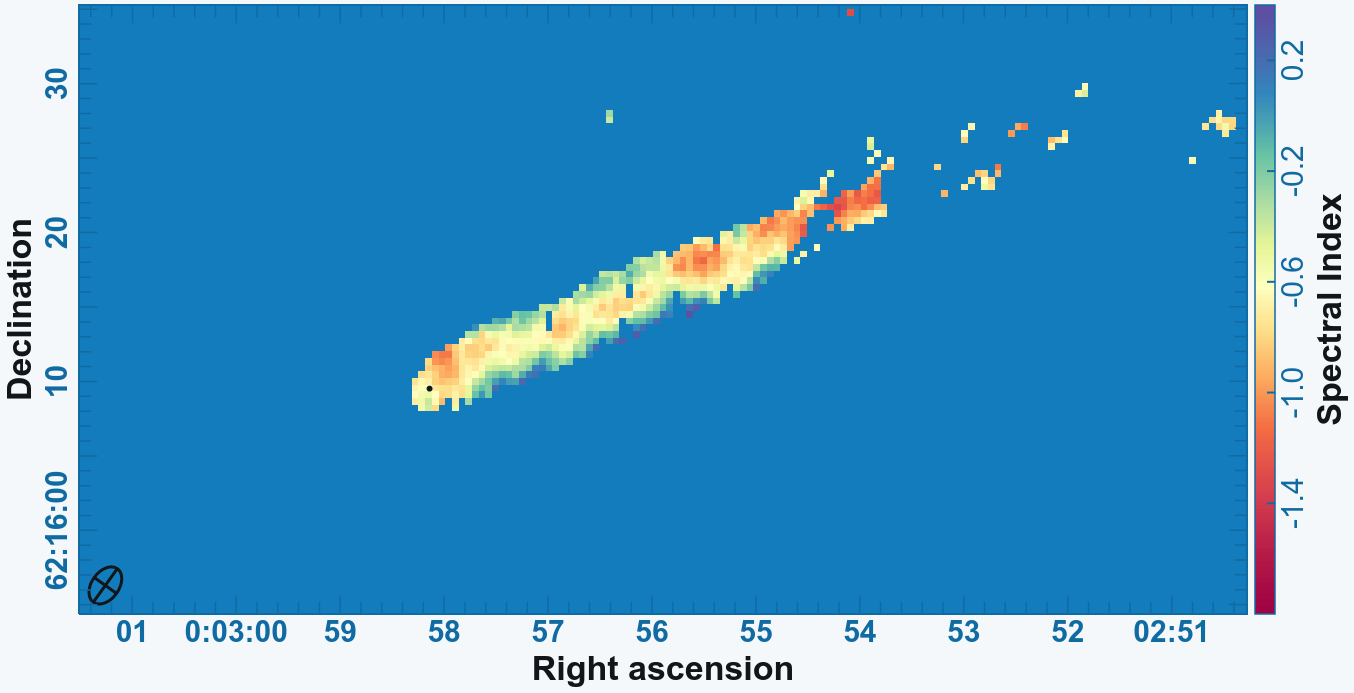}
    \caption{The spectral index image of the naturally weighted, combined 4--12\,GHz continuum image of the pulsar, with the location of the pulsar shown by the black dot, and the size of the VLA synthesized beam of 2.77$\arcsec$ x 1.84$\arcsec$ shown by the black ellipse in the lower left corner. Pixels below five times image noise rms and above 0.5 error in the spectral index are masked. It shows an inverted flat spectrum sheath (light green) along the tail of the PWN.}
    \label{fig:4}
\end{figure*}

\section{Results}\label{sec:results}

\begin{figure*}[htbp!]
    \centering
    \epsscale{0.8}
    \plotone{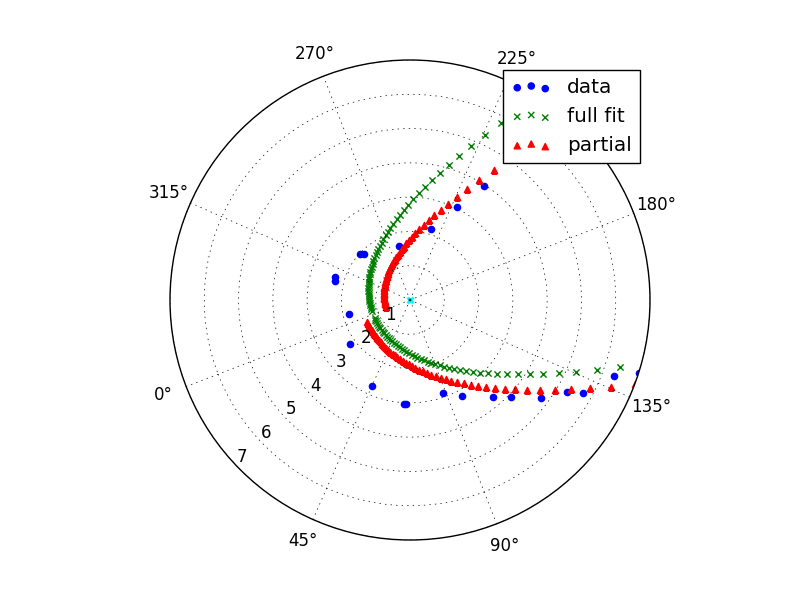}
    \caption{The figure shows the edge of emission, four times above the noise rms, 
    and centered on the pulsar (cyan), based on Figure \ref{fig:1} (blue). Fits are applied to measure the standoff distance (r$_s$) of the PWN using the analytical model in \citet{wilkins}. Green cross marks show a single fit applied to all the data and red triangles show the partial fits applied to each half of the emission separately, considering the asymmetry in emission. }
    \label{fig:8}
\end{figure*}

\begin{figure*}
    \centering
    \epsscale{1.3}
    \plotone{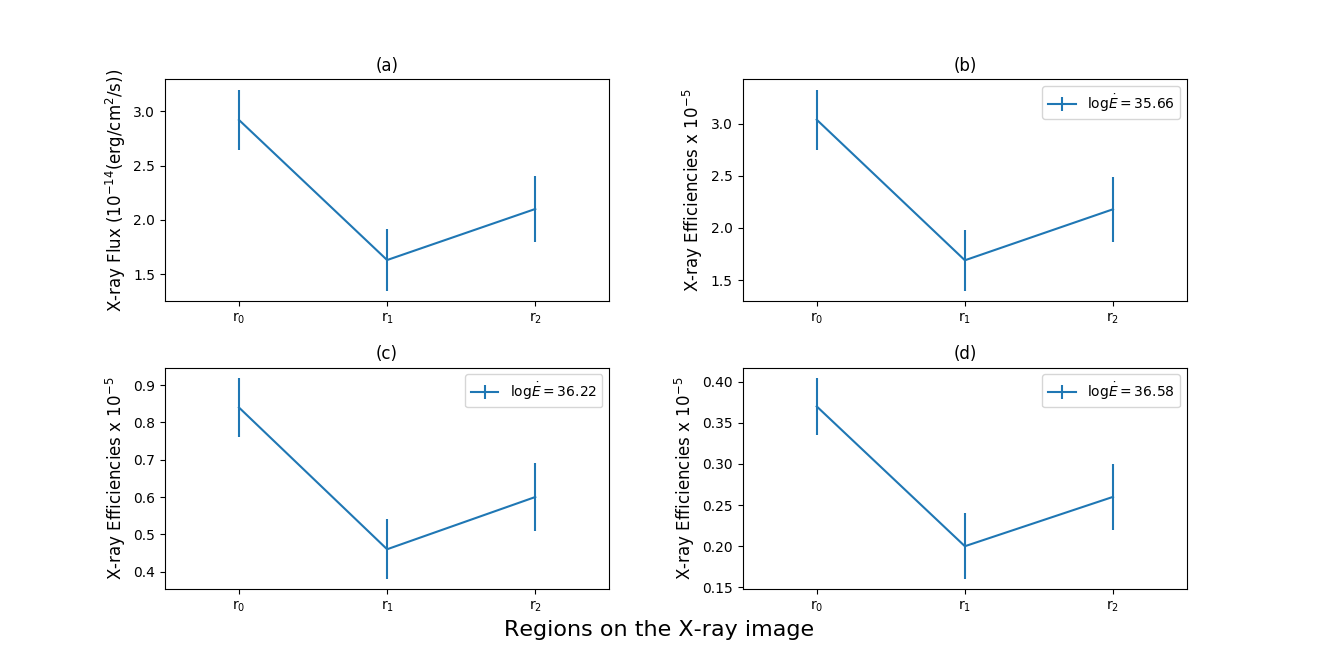}
    \caption{(a) Plot of X-ray Flux for the three regions marked in Figure \ref{fig:3}. (b,c,d) Plot of calculated X-ray efficiencies as a function of spatial bins in Figure \ref{fig:3}, for the three values of $log\dot{E}$ calculated in section \ref{sec:results}.}
    \label{fig:5}
\end{figure*}

\begin{figure*}[htbp!]
\gridline{\fig{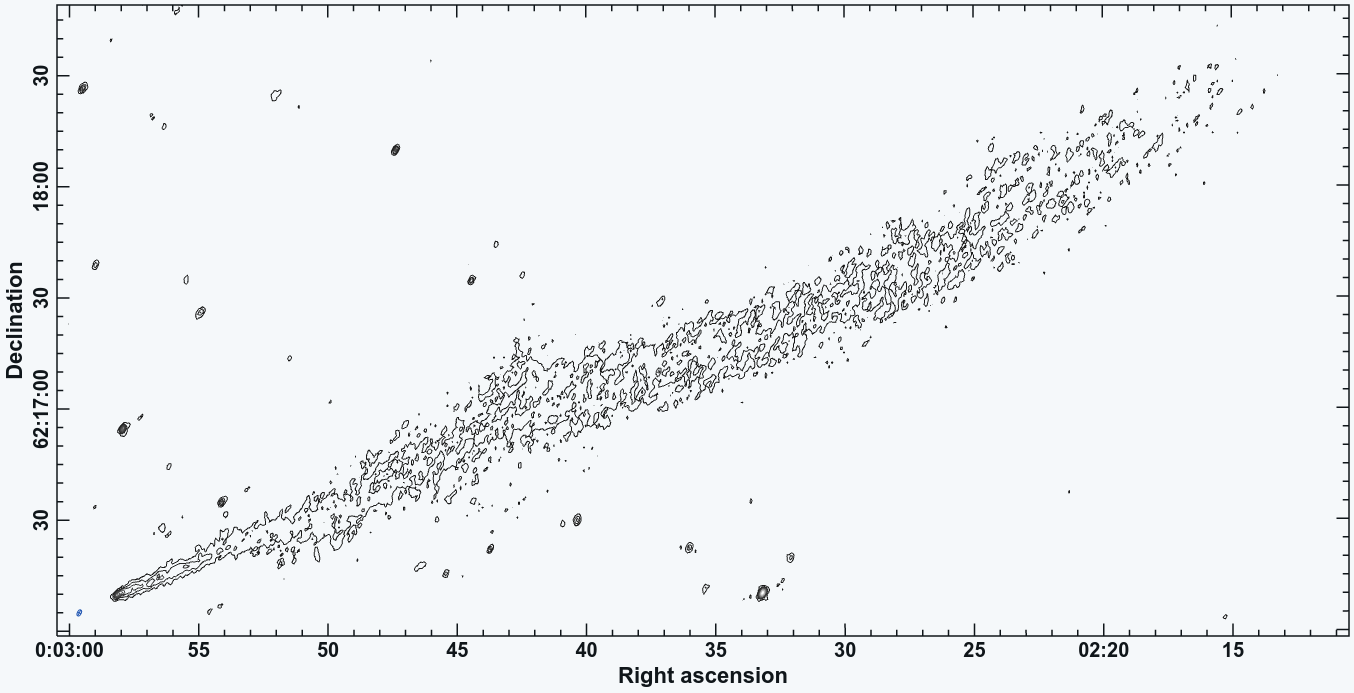}{0.8\textwidth}{(a)}}
\gridline{\fig{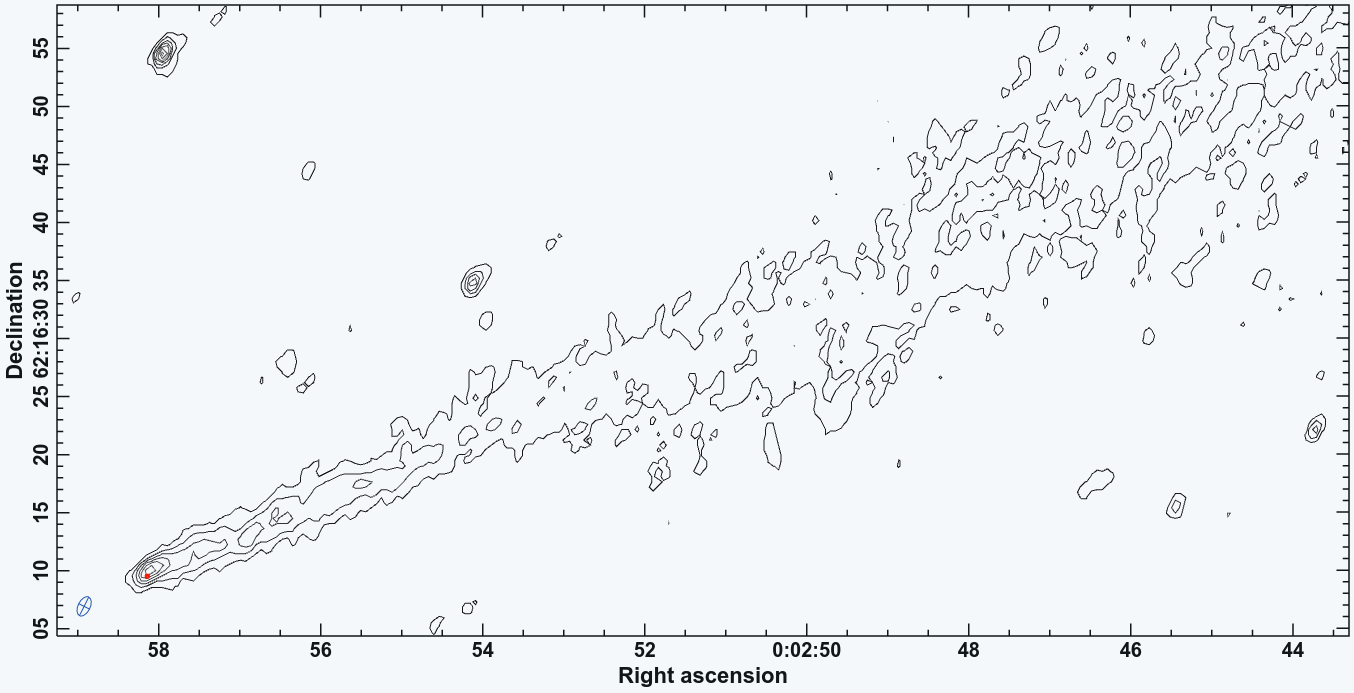}{0.8\textwidth}{(b)}}
\caption{Contour of the radio continuum image of the pulsar bow shock and the cometary tail between 4--12\,GHz. The contour levels are at 4,8,12,16,20, up to 76 sigma times the noise rms of 1.08 $\mu$Jy, in steps of 4 sigma. The size of the VLA synthesized beam of 1.78$\arcsec$ x 1.03$\arcsec$ is shown by the blue ellipse in the bottom left corner. (a) shows the extended emission behind the pulsar, along with two kinks in the cometary tail. (b) shows the zoom in on the bow shock region, with the position of the pulsar from \citet{frank} shown by the red dot, with positional error smaller than its size. The pulsar position is slightly shifted from the contour of highest emission.}
\label{fig:7}
\end{figure*}

Our resolved radio and X-ray image clearly shows the presence of a bow shock nebula as claimed by \citet{frank}. The combined broadband image in Figure \ref{fig:1} with ten times higher resolution than the previous 1--2\,GHz observations also shows a resolved and extended tail with scattered hot spots along it, which was only marginally resolved in \citet{frank} and the presence of two kinks, which was not clear in the 1--2\,GHz radio image. It presents an extended cometary tail $>$5.3$\arcmin$ extending up to the edge of the primary beam, with varying widths along the tail, thus in agreement with the size the of the tail seen in \citet{frank}. Unlike the uniformly bright emission seen in the low resolution 1--2\,GHz image from \citet{frank}, we see that the emission near the bow shock region and just behind the pulsar shows a highly collimated geometry extending to about $40\arcsec$, with the tail behind expanding laterally along the principal axis of emission, varying in brightness and with multiple hotspots. The location of the two kinks seen in Figures \ref{fig:1} and \ref{fig:7}, is consistent with the 1--2\,GHz continuum image in \citet{frank}, where the latter may suggest a possible third kink in the tail which is beyond the field of view of the current observations.\\
\vspace{-0.0cm}
The spectral index image, from the combined 4--12\,GHz C and X band data, from a naturally weighted image, shows a steep spectrum component. We measure a mean radio spectral index value of -0.56$\pm$0.36, defined as F$\propto$ ${\nu}^{\alpha}$, where F is the flux density, $\nu$ is the frequency and $\alpha$ is the radio spectral index, over the entire tail region of Figure \ref{fig:7}, still in agreement with -0.98$\pm$0.31 from \citet{frank}. Although the mean spectrum is not too steep, the figure also shows a steep spectrum channel of $\sim$ -1.2 to -1.4 along the inner region of the bow shock with with a flat spectrum sheath of $\sim$ -0.2 to 0.2 around it as shown by the red and light green colors in Figure \ref{fig:4}, each with uncertainty $<$0.5. While the two components were not resolved out in the previous 1--2\,GHz observation due to its limited angular resolution, the steep spectrum component is atypical of a PWN as suggested in \citet{weilersramek}. To look for evidence of spectral steepening along the tail we laterally average the points in the spectral index image from the principal axis of emission, but we do not see any clear evidence of steepening given uncertainties on each of those measurements.\\
\vspace{-0.0cm}
The X-ray emission fades quickly compared to the radio emission where the brightest feature near the pulsar extends only to 21$\arcsec$. This confirms the prediction of \citet{frank}, of an X-ray tail extending up to 7$\arcmin$ (t$_{\mathrm{syn}}$/t$_{\mathrm{R}}$) $\approx$ 20$\arcsec$, where t$_{\mathrm{R}}$ is the age of the radio PWN obtained simply by dividing its
angular size with the pulsar proper motion, t$_{\mathrm{syn}}$ is the synchrotron lifetime of X-ray electrons at 1\,keV and 7$\arcmin$ is the extent of radio tail at 1.5\, GHz. Based on our spectral analysis of different regions as shown in Figure \ref{fig:3}, we get the best fit and 90$\%$ confidence interval for the photon index parameter, whose values are given in Table \ref{tab:1}. The X-ray data shows marginal evidence for the evolution of the PWN emission along the tail, with spectral slope changes consistent with synchrotron cooling. The corresponding bounds calculated on the X-ray energy flux (F$_x$), in the energy range $0.5-8.0$\,keV, from each region are shown in Table \ref{tab:2} and the associated plot is shown in Figure \ref{fig:5}(a). From the X-ray flux we calculate the luminosity as L$_x$=4$\pi$d$^2$ F$_x$, where we use d=2\,kpc from \citet{frank}. This gives us the luminosity in the range of (5.65 - 8.33) $\times$ 10$^{30}$ ergs s$^{-1}$, (2.54 - 5.26) $\times$ 10$^{30}$ ergs s$^{-1}$ and (3.59 - 6.46) $\times$ 10$^{30}$ ergs s$^{-1}$ for the three regions $r_0$, $r_1$ and $r_2$ respectively, which is at least an order of magnitude lower than the typical values of 10$^{32}$ ergs s$^{-1}$ for such systems \citep[][]{kargalchandra}.\\
\vspace{-0.0cm}
To measure the position angle ($\theta_{\mu}$) of the pulsar, we apply cuts perpendicular to the symmetry axis of emission, for the bright tail near the pulsar position, in the combined radio image, to identify the brightest pixels in the transverse direction. We then apply a linear fit to measure $\theta_{\mu}$ = 111.13$\pm$0.52, which is in agreement with values derived in \citet{frank}, but with significantly improved constraints, thus confirming the geometry of the the system. To get a direct fit for the standoff distance (r$_s$) of the bow shock, following the model and using equation 9 of \citet{wilkins}\\
\begin{equation}
    R(\theta)=R_{o}\frac{\sqrt{3(1-\theta\cot\theta)}}{\sin\theta}
\end{equation}
where R and $\theta$ are the location of point on the bow shock in polar coordinates centered on the pulsar and R$_o$ is the standoff distance, we identify the edge of nebular emission near the pulsar as the point from the contour of emission, four times above the noise rms, in the combined radio image shown in Figure \ref{fig:1}. The points are then transformed to polar coordinates along the symmetry axis, centered on the pulsar, as shown by the blue dots in Figure \ref{fig:8}. Given the asymmetry in the bow shock emission, we consider the case of partial fits to each half of the emission, which gives us a standoff distance in the range of 0.36$\pm$0.05$\arcsec$ to 0.63$\pm$0.03$\arcsec$. Whereas, if we consider the distance between the tip of emission and the pulsar as the standoff distance, we get 0.95$\arcsec$. Considering the distance estimate of 2\,kpc in \citet{frank}, we get the standoff distance as 0.0032, 0.0061 and 0.0092\,pc respectively.\\
These values are among the lowest reported standoff distances and comparable to that of the Frying Pan nebula which has an estimated large Mach number M$\approx$\,200 \citep[][]{CNg}. \citet{kargalchandra} give R$_{TS,h}$ $\sim$ 0.04 $\dot{E}_{36}$ $^{1/2}$ n$^{-{1/2}}$(V$_p$/100\,kms$^{-1}$)$^{-1}$\,pc from \citet{kargalchandra}, where $\dot{E}$=10$^{36}$ $\dot{E}_{36}$\,ergs s$^{-1}$ is the pulsar spin-down energy, $n$ is the hydrogen column density (in units of 10$^{22}$\,cm$^{-2}$) and V$_p$ is the pulsar velocity. Using $n=0.595$ from HI4PI survey \citep[][]{h14pi} in the direction of the pulsar and V$_p$=1100\,kms$^{-1}$ from \citet{frank}, we get log$\dot{E}$ as, 35.66, 36.22 and 36.58\,ergs s$^{-1}$ respectively, for the three standoff distance estimates above. These values are similar to the value of $\dot{E}$ derived for other PWNs \citep[][]{kargalchandra}, however not all in agreement with log$\dot{E}$=35.18 from timing analysis \citep[][]{atnf}. The calculated value of $\dot{E}$ depends on the projected standoff distance based on the pulsar distance and the calculated velocity which in turn depends on the separation between the center of CTB1 and the tip of the pulsar and the pulsar proper motion, both of which are measured quantities from \citet{frank}, and the distance to the pulsar. Since the number density in the equation above is already taken from the HI4PI survey, the distance to the pulsar is the only adjustable parameter. Given the direct proportionality of V$_p$ and R$_{TS,h}$ on the pulsar distance, and similarly of $\dot{E}$ $^{1/2}$ on the product of the former two quantities, $\dot{E}$ depends directly on the distance to the pulsar. Hence, all three calculated values of $\dot{E}$ will not be in agreement with the timing value simultaneously. However, for an adjusted value lower than 2\,kpc all three can be in close agreement with the timing value, rather than the current scenario where all estimates are larger than the timing solution. Based on these measurements, we calculate the X-ray efficiency as $\eta_x$=L$_x$/$\dot{E}$, for each of the $\dot{E}$ values, for the three different pulsar regions, shown in Table \ref{tab:6} with corresponding plots shown in Figure \ref{fig:5}(b), \ref{fig:5}(c), and \ref{fig:5}(d). These values are of the order $10^{-5}$ and among the lowest seen in  X-ray PWNs \citep[][]{kargalchandra}.

\begin{deluxetable*}{cccc}[htbp!]
\tablecolumns{4}
\tabletypesize{\small}
\tablewidth{0pt}
\tablecaption{ X-ray Photon Index\label{tab:1}}
\tablehead{
    \colhead{Region } & \colhead{Best Fit} & \colhead{Lower Bound} & \colhead{Upper Bound} 
}
\startdata
r${_0}$& 4.12 & -1.59 & 4.72 \\ 
r${_1}$& 3.23 & -1.56 & 4.83 \\ 
r${_2}$& 5.69 & -2.48 & 6.91 \\ 
\enddata
\tablecomments{Best fit and 90$\%$ bounds for the photon index of the three regions listed above and in Figure \ref{fig:3}.}
\end{deluxetable*}

\begin{deluxetable*}{ccc}[htbp!]
\tablecolumns{3}
\tabletypesize{\small}
\tablewidth{0pt}
\tablecaption{ Derived X-ray Flux \label{tab:2}}
\tablehead{
    \colhead{Region } & \colhead{Lower Bound} & \colhead{Upper Bound} \\
    \colhead{} & \colhead{10$^{-14}$(erg/cm$^2$/s)} & \colhead{10$^{-14}$(erg/cm$^2$/s)}
}
\startdata
r${_0}$& 1.18 & 1.74 \\ 
r${_1}$& 0.53 & 1.10 \\ 
r${_2}$& 0.75 & 1.35 \\ 
\enddata
\tablecomments{Calculated 90$\%$ bounds for energy flux, in the $0.5-8.0$\,keV energy band, for the three regions listed above and shown in Figure \ref{fig:3}.}
\end{deluxetable*}

\begin{deluxetable*}{cccc}[htbp!]
\tablecolumns{4}
\tabletypesize{\small}
\tablewidth{0pt}
\tablecaption{ Calculated X-ray Efficiencies\label{tab:6}}
\tablehead{
    \colhead{} & \colhead{Region r$_0$} & \colhead{Region r$_1$} & \colhead{Region r$_2$}\\
    \colhead{} & \colhead{(10$^{-5}$)} & \colhead{(10$^{-5}$)} & \colhead{(10$^{-5}$)}
}
\startdata
log$\dot{E}$=35.66 & 1.23-1.81 & 0.55-1.14  & 0.78-1.40 \\ 
log$\dot{E}$=36.22 & 0.34-0.5 & 0.15-0.31 & 0.21-0.39 \\ 
log$\dot{E}$=36.58 & 0.15-0.22 & 0.06-0.14 & 0.09-0.17 \\ 
\enddata
\tablecomments{Table of calculated X-ray efficiencies in the order of 10$^{-5}$, for the three different spatial regions, for each of the three different cases of calculate $\dot{E}$ values.}
\end{deluxetable*}

\vspace{-2.5cm}

\section{Discussion} \label{sec:discuss}
\subsection{The radio and X-ray tail}\label{sec:d1}
The radio images of the PWN show a number of remarkable features, including the clear shape of the bow shock at the tip of emission, and an elongated cometary tail in the image; at ten times higher resolution from the previous 1--2\,GHz observations in \citet{frank}. Similar to the Frying pan nebula \citep[][]{CNg} the width of the elongated tail appears to expand from 3$\arcsec$ to 23$\arcsec$ as shown in Figure \ref{fig:1}, similar to the former where expansion by a factor of 10 is seen in the tail. Since both systems have similar Mach numbers of $\approx$ 200, leading to a very sharp Mach cone, we attribute this expansion in the tail to overpressure, as shown by \citet{CNg}. The presence of hotspots in the extended radio tail, far from the pulsar, indicates the role of ISM turbulence thereby leading to compressions and enhancing synchrotron emission from these locations. The presence of turbulence can be verified via the deviation of magnetic field topology from nearby regions. At the same time we cannot ignore density fluctuations. An enhanced electron density under similar magnetic field conditions can also explain the presence of these feature in the radio image. On the other hand X-ray emission extends up to 21$\arcsec$, and becomes more collimated with increasing distance from the pulsar, as shown in Figure \ref{fig:3}. The case of X-rays is similar to the case of the ``Mouse" \citep[][]{mouse}, however in the latter the radio emission just behind the pulsar shows expansion, and here we see a uniformly bright emission up to $\approx$ 10-15$\arcsec$. The narrowing of the X-ray feature could be explained by more efficient cooling of X-ray emitting electrons away from the symmetry axis of emission, either due to a stronger magnetic field, or a higher velocity of flow in the inner channel of the tail. For the former to occur, the synchrotron emission should be enhanced in both the X-ray and radio in the tail region, unless the electron density decreases in proportion. We see that the X-ray emission diminishes and although the radio remains uniformly bright, given the longer cooling time of radio emitting electrons, they may not be affected on similar distances as the X-ray nebula, rendering both radio and X-ray emission in deficit of synchrotron enhancement. Hence, this seems an unlikely scenario. The other case requires the flow velocity to increases towards the symmetry axis from the edge of nebular emission. However this is in contrast to the simulation presented in \citealt{buccilaminar}, where the author shows a slower inner channel of flow; although, this may be the result of limitations in the simulation setup. We also see a break in emission within the tail of the X-ray data, which was also present in the combined 8--12\,GHz X band radio image \footnote{\label{note1}image not shown here because of redundancy} of the pulsar, whose origin is not clear. Apart from this, the X-ray tail fades away quickly, with a few hot spots scattered along the extended cometary tail, following the radio emission, evident as the scattered emission inside the cyan radio contour in Figure \ref{fig:3}.
\subsection{The radio and X-ray spectrum}\label{sec:d2}
The spectral index image shown in Figure \ref{fig:4} does not show explicit evidence of spectral steepening with distance from the pulsar. We see the elongated tail fading out quickly in our 8--12\,GHz X band image compared to the 4--8\,GHz C band image \footnoteref{note1}. Nevertheless, the radio PWN spectral index value in general has a flatter spectrum in the range of -\,0.5 to 0 as suggested by \citet{weilersramek} based on a survey of supernova remnants conducted in the frequency range of 100\, MHz to 10\, GHz.  We do see a much steeper spectrum channel down the middle of the emission region of $\sim$\,-1.2\, to \,-1.4 in Figure \ref{fig:4}, and close to the bow shock region.  We find a flat spectrum sheath towards the elongated tail closer to the PWN-ISM boundary, from the combined broadband data, in better agreement with the expected range above. It is worth pointing out that we were not able to obtain spectral images with good confidence level from the individually combined 8--12\,GHz X band or 4--8\,GHz C band data. Since some of these multi-term multi-frequency imaging routines are more recently developed, the authors here, are not aware of any other high resolution spectral images of such systems. Spectral flattening towards the edges may be related to loss of energy due to collisions at the surface or injection of energy at higher frequency from reverse shocks. Another possibility is the decrease in opacity in the inner channel of the nebula, due to the shock fronts traveling outwards. Since this region is close the pulsar and appears uniformly bright in Figure \ref{fig:1}, turbulence in the medium seems unlikely. Based on the simulations presented in \citealt{buccilaminar}, Doppler boosting could also account for this. Since the simulations suggest a slower inner channel of flow the emission in the inner region will be less de-boosted compared to the surrounding and will appear brighter, thereby steeper on the inside.\\
The X-ray data provides marginal evidence for the evolution of the PWN emission along the tail, with spectral slope changes consistent with synchrotron cooling \citep[][]{broadsynchro}. Generally PWN with soft spectra show higher X-ray efficiencies \citep[][]{kargalchandra}, however in this case we see that, although the X-ray spectrum is soft the efficiency is among the lowest reported for such systems. This could be slightly improved if the pulsar was farther than 2\,kpc, which was not ruled out in \citet{frank}. However, that would lead to disagreement between our measured $\dot{E}$ and that from the timing analysis.\\
\subsection{Anomalous features: kinks, laminar flows, asymmetric emission}\label{sec:d3}
We also see multiple kinks in the extended cometary tail in the combined broad-band image in Figure \ref{fig:1}. A kink is also seen in the radio image of the Frying pan nebula, which does not have an associated X-ray emission \citep[][]{CNg}.  The dynamics of the system are driven by the magnetic field from the pulsar and its interaction with the ambient field of the ISM (which is much weaker), the density variations or gradients in the local ISM, the velocity of the pulsar, and its orientation with respect to density structures and the inclination angle with spin axis as well as with the magnetic field direction. Previous hydrodynamic and non-relativistic studies under some assumptions of symmetry \citep[][]{toropina,yoondoo} suggest the role of density discontinuities and orientation of the system. A full 3D relativistic magneto-hydrodynamic (MHD) treatment of the problem \citep[][]{olmibuccidynamics}, considering all parameters, also shows that density discontinuities in the ISM could be responsible for such features in the emission of these objects. But since the tail resumes its original orientation further down the kink, it might be difficult to do so via density gradients. Other possibilities include flow instability and ISM turbulence. In the former case we expect a disorder in the magnetic field while the latter requires a high flow speed to compress the post-shock wind and would also lead to reduced polarization near the kinks. For the case of the Frying pan nebula the former argument was rejected based on the magnetic field orientation which remained ordered across the kink and the tail also resumed its original orientation \citep[][]{CNg}. The tail resumes its original orientation in our case as well, however we need the magnetic field orientation to further explore these possibilities. As for the case of ISM turbulence, it will lead to enhanced radio emission near the kinks as was seen in the case of the Frying pan nebula. However, there is not sufficient evidence of enhancement in our case to suggest an origin via ISM turbulence.\\
\vspace{-0.0cm}
Figure \ref{fig:7}b shows the broad-band radio contour, zooming in on the emission region near the pulsar and the head of the bow shock. The emission contours show a laminar structure with a central cone of emission behind the pulsar which spreads out in a stepped fashion towards the PWN-ISM boundary. A similar structure is also present in the spectral index image of Figure \ref{fig:4}, which shows a central steep spectrum component. This could be indicative of a laminar magnetic field structure behind the pulsar, as pointed out earlier by \citet{buccilaminar}, where simulations show that the flow speed varies orthogonally to the symmetry axis of emission, with a slowly moving inner channel.\\
\vspace{-0.0cm}
As pointed out earlier in section \ref{sec:results}, we see asymmetrical emission in the bow shock region, which is evident from both the contour of X-ray emission as well as the asymmetrical emission present in the highest resolution 8--12\,GHz image of the pulsar in Figure \ref{fig:2}. Again, such features could arise out of strong turbulence in the region leading to compression and enhancement of synchrotron emissivity, but may also be indicative of jet like features originating from the pulsar \citep[see, e.g.,][]{devries}. Such jet like features arise due to the escape of high energy particles from the PWN along the magnetic field lines of the ISM \citep[][]{olmiescape} or could be jets from the equatorial region where the magnetic field is stronger. Hence, it's possible that the asymmetric hot spots of emission we see near the pulsar in the radio image shown in Figure \ref{fig:2} is one such jet. At the same time, we cannot ignore the role of relative orientation of the magnetic field, pulsar velocity (inclination) and the observer direction with respect to the plane of the sky oriented along pulsar velocity. As demonstrated by \citet{buccilaminar}, the simulations show one sided asymmetric emission in total intensity images at the tip of nebula only when both the inclination angle and the observer direction are near 45\,\degr, which cannot be ruled out in our case.\\
Comparing the values of $\dot{E}$ derived in section \ref{sec:results}, with the value reported in the ATNF pulsar catalog \citep[][]{atnf} for this pulsar (log$\dot{E}$=35.18) from the timing analysis, our lower value of 35.66 is in reasonable agreement, while the other two estimates from section \ref{sec:results} are about an order of magnitude higher.  The first two values would be in closer agreement if the distance estimate is smaller by a factor of $\sim$2 or more. But in neither of these cases does a model of the bow shock nebula completely trace the emission from the radio images, as shown clearly in Figure \ref{fig:8}, leaving some excess emission in front and towards the side of the pulsar. This could be another indication that jets are present in the system. Although we don't see any clear indication of a jet like feature in the X-ray image unlike for some other PWNs, the asymmetry is present as indicated by the X-ray emission contour in Figure \ref{fig:2}. This is likely due to the very high velocity of the pulsar, making the jets bend and collimate behind the pulsar. If dual jets exist, then the system must have a high inclination angle so that only one of the jets appears as a hot spot while the other gets dimmer due to Doppler effects. However, this would also suggest that our measured standoff distances are smaller than expected due to projection effects which would add to the deviation between our measured $\dot{E}$ value and those obtained from the timing analysis.\\ 

\vspace{-0.1cm}
\section{Summary and Conclusions}\label{sec:summary}
We have presented here a multi-frequency study of the bow shock and extended tail of emission of the recently discovered PWN of PSR J0002+6216. We have carried out a multi-epoch follow-up campaign to study the bow shock region of this system since its initial discovery and association with the supernova remnant CTB1 \citep[][]{frank}. Our radio images at higher frequencies show an extended tail, consistent with the previous findings at the lower frequency 1.5\,GHz observations. A fit to the bright emission near the pulsar is used to provide an improved measurement of the position angle of the pulsar. Additionally, we see multiple kinks along the cometary tail, which point towards density variations in the ISM or possibly turbulence. A hint of such structures is also present in the lower frequency radio observations. The spectral index image obtained from the combined broadband data shows an unusually steep spectrum channel, with a flat spectrum sheath present around it. This could be a result of lower opacity in the inner part of the cometary tail or due to the injection of energy at higher frequencies along the edge of the tail, near the shock fronts.\\
\vspace{-0.0cm}
We use the emission near the pulsar from the combined broadband image to constrain the bow shock geometry and obtain an estimate of the standoff distance. Our findings show a very small bow shock region, with corresponding large spin down energy, in reasonable agreement with that obtained from timing solutions of the pulsar. Our highest resolution radio image also reveals asymmetry in the bow shock region, showing misaligned hot spots from the location of the pulsar. Such features could occur due to the presence of density discontinuities, turbulence or escape of high energy particles from the region or due to jets originating from the pulsar itself. In the latter case, the system would require a high inclination angle to suppress the emission of the other jet via Doppler effects.\\
\vspace{-0.0cm}
Finally, we have also done an X-ray study of the system using spatially resolved observation from CXO. These provide marginal evidence of cooling along the tail. Although jets are not distinctly present in these observations, the presence of an asymmetry in the bow shock is consistent with the radio emission. This could indicate collimation of these jets in the direction of the cometary tail due to the very high velocity of the pulsar. The X-ray spectra also reveals an unusually low X-ray efficiency for the system.\\
\vspace{-0.0cm}
Additional X-ray observations are required to fully characterize the particle spectrum as well as to understand the nature of asymmetry in the bow shock region. Also, polarization measurements at radio wavelengths can reveal the structure of the magnetic field inside the nebula and the nearby region to better understand the dynamics of the system, as well as to better characterize the various unusual features seen in the radio images.
\vspace{-0.2cm}
\section{Acknowledgements}
We thank the referee for constructive suggestions. The authors thank Dale A. Frail for helpful discussions and suggestions. The National Radio Astronomy Observatory is a facility of the National Science Foundation operated under cooperative agreement by Associated Universities, Inc. Support for this work was provided by the National Aeronautics and Space Administration through Chandra Award Number GO0-21051X issued by the Chandra X-ray Center, which is operated by the Smithsonian Astrophysical Observatory for and on behalf of the National Aeronautics Space Administration under contract NAS8-03060.  Work at NRL is supported by NASA.

\facilities{VLA, Chandra}
\software{CASA \citep[][]{casa}, CARTA \citep[][]{comriecarta}, CIAO \citep[][]{ciao}, Numpy \citep[][]{npy}, Scipy \citep[][]{scipy}, matplotlib \citep[][]{matplot}}


\bibliographystyle{aasjournal}
\bibliography{cannonballbib}

\begin{thebibliography}{}
\expandafter\ifx\csname natexlab\endcsname\relax\def\natexlab#1{#1}\fi
\providecommand{\url}[1]{\href{#1}{#1}}
\providecommand{\dodoi}[1]{doi:~\href{http://doi.org/#1}{\nolinkurl{#1}}}
\providecommand{\doeprint}[1]{\href{http://ascl.net/#1}{\nolinkurl{http://ascl.net/#1}}}
\providecommand{\doarXiv}[1]{\href{https://arxiv.org/abs/#1}{\nolinkurl{https://arxiv.org/abs/#1}}}

\bibitem[{{Bhatnagar} {et~al.}(2013){Bhatnagar}, {Rau}, \& {Golap}}]{bhatnagar}
{Bhatnagar}, S., {Rau}, U., \& {Golap}, K. 2013, \apj, 770, 91,
  \dodoi{10.1088/0004-637X/770/2/91}

\bibitem[{{Briggs}(1995)}]{danbriggs}
{Briggs}, D.~S. 1995, in American Astronomical Society Meeting Abstracts, Vol.
  187, American Astronomical Society Meeting Abstracts, 112.02

\bibitem[{{Bucciantini}(2018{\natexlab{a}})}]{bucchiantini2018}
{Bucciantini}, N. 2018{\natexlab{a}}, \mnras, 480, 5419,
  \dodoi{10.1093/mnras/sty2237}

\bibitem[{{Bucciantini}(2018{\natexlab{b}})}]{buccilaminar}
---. 2018{\natexlab{b}}, \mnras, 478, 2074, \dodoi{10.1093/mnras/sty1199}

\bibitem[{{Bykov} {et~al.}(2017){Bykov}, {Amato}, {Petrov}, {Krassilchtchikov},
  \& {Levenfish}}]{bykov}
{Bykov}, A.~M., {Amato}, E., {Petrov}, A.~E., {Krassilchtchikov}, A.~M., \&
  {Levenfish}, K.~P. 2017, \ssr, 207, 235, \dodoi{10.1007/s11214-017-0371-7}

\bibitem[{Chatterjee \& Cordes(2003)}]{guitar}
Chatterjee, S., \& Cordes, J.~M. 2003, The Astrophysical Journal, 600, L51,
  \dodoi{10.1086/381498}

\bibitem[{{Comrie} {et~al.}(2021){Comrie}, {Wang}, {Hsu}, {Moraghan}, {Harris},
  {Pang}, {Pi{\'n}ska}, {Chiang}, {Chang}, {Hwang}, {Jan}, {Lin}, \&
  {Simmonds}}]{comriecarta}
{Comrie}, A., {Wang}, K.-S., {Hsu}, S.-C., {et~al.} 2021, {CARTA: The Cube
  Analysis and Rendering Tool for Astronomy}, 2.0.0, Zenodo,  Zenodo,
  \dodoi{10.5281/zenodo.3377984}

\bibitem[{{de Vries} \& {Romani}(2022)}]{devries}
{de Vries}, M., \& {Romani}, R.~W. 2022, \apj, 928, 39,
  \dodoi{10.3847/1538-4357/ac5739}

\bibitem[{{Fruscione} {et~al.}(2006){Fruscione}, {McDowell}, {Allen},
  {Brickhouse}, {Burke}, {Davis}, {Durham}, {Elvis}, {Galle}, {Harris},
  {Huenemoerder}, {Houck}, {Ishibashi}, {Karovska}, {Nicastro}, {Noble},
  {Nowak}, {Primini}, {Siemiginowska}, {Smith}, \& {Wise}}]{ciao}
{Fruscione}, A., {McDowell}, J.~C., {Allen}, G.~E., {et~al.} 2006, in Society
  of Photo-Optical Instrumentation Engineers (SPIE) Conference Series, Vol.
  6270, Society of Photo-Optical Instrumentation Engineers (SPIE) Conference
  Series, ed. D.~R. {Silva} \& R.~E. {Doxsey}, 62701V,
  \dodoi{10.1117/12.671760}

\bibitem[{{Gelfand} {et~al.}(2009){Gelfand}, {Slane}, \& {Zhang}}]{gelfand}
{Gelfand}, J.~D., {Slane}, P.~O., \& {Zhang}, W. 2009, \apj, 703, 2051,
  \dodoi{10.1088/0004-637X/703/2/2051}

\bibitem[{Harris {et~al.}(2020)Harris, Millman, van~der Walt, Gommers,
  Virtanen, Cournapeau, Wieser, Taylor, Berg, Smith, Kern, Picus, Hoyer, van
  Kerkwijk, Brett, Haldane, del R{\'{i}}o, Wiebe, Peterson,
  G{\'{e}}rard-Marchant, Sheppard, Reddy, Weckesser, Abbasi, Gohlke, \&
  Oliphant}]{npy}
Harris, C.~R., Millman, K.~J., van~der Walt, S.~J., {et~al.} 2020, Nature, 585,
  357, \dodoi{10.1038/s41586-020-2649-2}

\bibitem[{{HI4PI Collaboration} {et~al.}(2016){HI4PI Collaboration}, {Ben
  Bekhti}, {Fl{\"o}er}, {Keller}, {Kerp}, {Lenz}, {Winkel}, {Bailin},
  {Calabretta}, {Dedes}, {Ford}, {Gibson}, {Haud}, {Janowiecki}, {Kalberla},
  {Lockman}, {McClure-Griffiths}, {Murphy}, {Nakanishi}, {Pisano}, \&
  {Staveley-Smith}}]{h14pi}
{HI4PI Collaboration}, {Ben Bekhti}, N., {Fl{\"o}er}, L., {et~al.} 2016, \aap,
  594, A116, \dodoi{10.1051/0004-6361/201629178}

\bibitem[{Hobbs {et~al.}(2005)Hobbs, Lorimer, Lyne, \& Kramer}]{hobbs}
Hobbs, G., Lorimer, D.~R., Lyne, A.~G., \& Kramer, M. 2005, Monthly Notices of
  the Royal Astronomical Society, 360, 974,
  \dodoi{10.1111/j.1365-2966.2005.09087.x}

\bibitem[{{Hooper} \& {Linden}(2018)}]{Hooper}
{Hooper}, D., \& {Linden}, T. 2018, \prd, 98, 043005,
  \dodoi{10.1103/PhysRevD.98.043005}

\bibitem[{Hunter(2007)}]{matplot}
Hunter, J.~D. 2007, Computing in Science \& Engineering, 9, 90,
  \dodoi{10.1109/MCSE.2007.55}

\bibitem[{Jones {et~al.}(2001)Jones, Oliphant, \& Peterson}]{scipy}
Jones, E.~D., Oliphant, T.~E., \& Peterson, P. 2001

\bibitem[{{Joye} \& {Mandel}(2003)}]{DS9}
{Joye}, W.~A., \& {Mandel}, E. 2003, in Astronomical Society of the Pacific
  Conference Series, Vol. 295, Astronomical Data Analysis Software and Systems
  XII, ed. H.~E. {Payne}, R.~I. {Jedrzejewski}, \& R.~N. {Hook}, 489

\bibitem[{{Kargaltsev} \& {Pavlov}(2008)}]{kargalchandra}
{Kargaltsev}, O., \& {Pavlov}, G.~G. 2008, in American Institute of Physics
  Conference Series, Vol. 983, 40 Years of Pulsars: Millisecond Pulsars,
  Magnetars and More, ed. C.~{Bassa}, Z.~{Wang}, A.~{Cumming}, \& V.~M.
  {Kaspi}, 171--185, \dodoi{10.1063/1.2900138}

\bibitem[{{Kargaltsev} {et~al.}(2017){Kargaltsev}, {Pavlov}, {Klingler}, \&
  {Rangelov}}]{kargaltsevklinger}
{Kargaltsev}, O., {Pavlov}, G.~G., {Klingler}, N., \& {Rangelov}, B. 2017,
  Journal of Plasma Physics, 83, 635830501, \dodoi{10.1017/S0022377817000630}

\bibitem[{{Kennel} \& {Coroniti}(1984)}]{KennelCoroniti}
{Kennel}, C.~F., \& {Coroniti}, F.~V. 1984, \apj, 283, 694,
  \dodoi{10.1086/162356}

\bibitem[{{Klingler} {et~al.}(2018){Klingler}, {Kargaltsev}, {Pavlov}, {Ng},
  {Beniamini}, \& {Volkov}}]{mouse}
{Klingler}, N., {Kargaltsev}, O., {Pavlov}, G.~G., {et~al.} 2018, \apj, 861, 5,
  \dodoi{10.3847/1538-4357/aac6e010.48550/arXiv.1803.10294}

\bibitem[{{Manchester} {et~al.}(2005){Manchester}, {Hobbs}, {Teoh}, \&
  {Hobbs}}]{atnf}
{Manchester}, R.~N., {Hobbs}, G.~B., {Teoh}, A., \& {Hobbs}, M. 2005, \aj, 129,
  1993

\bibitem[{{Mohan} \& {Rafferty}(2015)}]{PyBDSFPB}
{Mohan}, N., \& {Rafferty}, D. 2015, {PyBDSF: Python Blob Detection and Source
  Finder}, Astrophysics Source Code Library, record ascl:1502.007.
\newblock \doeprint{1502.007}

\bibitem[{Morlino {et~al.}(2015)Morlino, Lyutikov, \& Vorster}]{morlino}
Morlino, G., Lyutikov, M., \& Vorster, M. 2015, Monthly Notices of the Royal
  Astronomical Society, 454, 3886, \dodoi{10.1093/mnras/stv2189}

\bibitem[{{Ng} {et~al.}(2012){Ng}, {Bucciantini}, {Gaensler}, {Camilo},
  {Chatterjee}, \& {Bouchard}}]{CNg}
{Ng}, C.~Y., {Bucciantini}, N., {Gaensler}, B.~M., {et~al.} 2012, \apj, 746,
  105, \dodoi{10.1088/0004-637X/746/1/105}

\bibitem[{{Olmi} \& {Bucciantini}(2019)}]{olmibuccidynamics}
{Olmi}, B., \& {Bucciantini}, N. 2019, \mnras, 484, 5755,
  \dodoi{10.1093/mnras/stz382}

\bibitem[{{Olmi} \& {Bucciantini}(2020)}]{olmiescape}
---. 2020, \memsai, 91, 321

\bibitem[{Perley {et~al.}(2009)Perley, Napier, Jackson, Butler, Carlson, Fort,
  Dewdney, Clark, Hayward, Durand, Revnell, \& McKinnon}]{perley}
Perley, R., Napier, P., Jackson, J., {et~al.} 2009, Proceedings of the IEEE,
  97, 1448, \dodoi{10.1109/JPROC.2009.2015470}

\bibitem[{{Rau} {et~al.}(2016){Rau}, {Bhatnagar}, \& {Owen}}]{Raubhatnagar}
{Rau}, U., {Bhatnagar}, S., \& {Owen}, F.~N. 2016, \aj, 152, 124,
  \dodoi{10.3847/0004-6256/152/5/124}

\bibitem[{{Rau} \& {Cornwell}(2011)}]{Raucornwell}
{Rau}, U., \& {Cornwell}, T.~J. 2011, \aap, 532, A71,
  \dodoi{10.1051/0004-6361/201117104}

\bibitem[{{Reynolds} \& {Chevalier}(1984)}]{Reynolds}
{Reynolds}, S.~P., \& {Chevalier}, R.~A. 1984, \apj, 278, 630,
  \dodoi{10.1086/161831}

\bibitem[{{Schinzel} {et~al.}(2019){Schinzel}, {Kerr}, {Rau}, {Bhatnagar}, \&
  {Frail}}]{frank}
{Schinzel}, F.~K., {Kerr}, M., {Rau}, U., {Bhatnagar}, S., \& {Frail}, D.~A.
  2019, \apjl, 876, L17, \dodoi{10.3847/2041-8213/ab18f7}

\bibitem[{Slane(2017)}]{slanepwn}
Slane, P. 2017, Pulsar Wind Nebulae (Cham: Springer International Publishing),
  2159--2179, \dodoi{10.1007/978-3-319-21846-5_95}

\bibitem[{{THE CASA TEAM} {et~al.}(2022){THE CASA TEAM}, {Bean}, {Bhatnagar},
  {Castro}, {Donovan Meyer}, {Emonts}, {Garcia}, {Garwood}, {Golap}, {Gonzalez
  Villalba}, {Harris}, {Hayashi}, {Hoskins}, {Hsieh}, {Jagannathan},
  {Kawasaki}, {Keimpema}, {Kettenis}, {Lopez}, {Marvil}, {Masters},
  {McNichols}, {Mehringer}, {Miel}, {Moellenbrock}, {Montesino}, {Nakazato},
  {Ott}, {Petry}, {Pokorny}, {Raba}, {Rau}, {Schiebel}, {Schweighart},
  {Sekhar}, {Shimada}, {Small}, {Steeb}, {Sugimoto}, {Suoranta}, {Tsutsumi},
  {van Bemmel}, {Verkouter}, {Wells}, {Xiong}, {Szomoru}, {Griffith},
  {Glendenning}, \& {Kern}}]{casa}
{THE CASA TEAM}, {Bean}, B., {Bhatnagar}, S., {et~al.} 2022, arXiv e-prints,
  arXiv:2210.02276.
\newblock \doarXiv{2210.02276}

\bibitem[{Toropina {et~al.}(2019)Toropina, Romanova, \& Lovelace}]{toropina}
Toropina, O.~D., Romanova, M.~M., \& Lovelace, R. V.~E. 2019, Monthly Notices
  of the Royal Astronomical Society, 484, 1475, \dodoi{10.1093/mnras/stz034}

\bibitem[{{van der Swaluw}(2003)}]{vandermhdpwn}
{van der Swaluw}, E. 2003, \aap, 404, 939, \dodoi{10.1051/0004-6361:20030452}

\bibitem[{{Weiler} \& {Sramek}(1988)}]{weilersramek}
{Weiler}, K.~W., \& {Sramek}, R.~A. 1988, \araa, 26, 295,
  \dodoi{10.1146/annurev.aa.26.090188.001455}

\bibitem[{{Wilkin}(1996)}]{wilkins}
{Wilkin}, F.~P. 1996, \apjl, 459, L31, \dodoi{10.1086/309939}

\bibitem[{{Xu} {et~al.}(2019){Xu}, {Klingler}, {Kargaltsev}, \&
  {Zhang}}]{broadsynchro}
{Xu}, S., {Klingler}, N., {Kargaltsev}, O., \& {Zhang}, B. 2019, \apj, 872, 10,
  \dodoi{10.3847/1538-4357/aafb2e}

\bibitem[{Yoon \& Heinz(2016)}]{yoondoo}
Yoon, D., \& Heinz, S. 2016, Monthly Notices of the Royal Astronomical Society,
  464, 3297, \dodoi{10.1093/mnras/stw2590}

\bibitem[{{Zhang} {et~al.}(2008){Zhang}, {Chen}, \& {Fang}}]{zhang}
{Zhang}, L., {Chen}, S.~B., \& {Fang}, J. 2008, \apj, 676, 1210,
  \dodoi{10.1086/527466}

\end{thebibliography}





\end{document}